\documentclass[a4paper,11pt]{article}

\usepackage{amsmath}
\usepackage{amssymb}
\usepackage{amsthm}
\usepackage{indentfirst}
\usepackage{graphicx}
\usepackage[utf8]{inputenc}
\usepackage[dvipsnames]{xcolor}
\usepackage{tabularx}
\usepackage{caption}
\usepackage{subcaption}
\usepackage{gensymb}
\usepackage[normalem]{ulem}

\usepackage{jheppub}
\hypersetup{colorlinks=true, linkcolor=blue, urlcolor=blue, citecolor=blue, linktocpage=true}

\begin{document}

\title{Deep learning method for identifying mass composition of ultra-high-energy cosmic rays}

\author[a,b] {O. Kalashev,}
\author[a,b] {I. Kharuk,}
\author[a,c] {M. Kuznetsov,}
\author[a] {G. Rubtsov,}
\author[d] {T. Sako,}
\author[e,f] {Y. Tsunesada,}
\author[a,d] {Ya. Zhezher.}

\affiliation[a]{Institute for Nuclear Research of the Russian Academy of Sciences,
\\ 60th October Anniversary Prospect, 7a, Moscow, 117312, Russia}
\affiliation[b]{Moscow Institute of Physics and Technology,\\
Institutsky lane 9, Dolgoprudny, Moscow region, 141700, Russia}
\affiliation[c]{Service de Physique Théorique, Université Libre de Bruxelles,\\
Boulevard du Triomphe CP225, 1050 Brussels, Belgium}
\affiliation[d]{Institute for Cosmic Ray Research (ICRR), University of Tokyo\\
5-1-5 Kashiwanoha, Kashiwa-shi, Chiba 277-8582, Japan}
\affiliation[e]{Graduate School of Science, Osaka City University, \\ Osaka, Osaka, 558-8585, Japan}
\affiliation[f]{Nambu Yoichiro Institute of Theoretical and Experimental Physics, Osaka City University, \\ Osaka, Osaka, 558-8585, Japan}

\emailAdd{ivan.kharuk@phystech.edu}

\abstract{We introduce a novel method for identifying the mass composition of ultra-high-energy cosmic rays using deep learning. The key idea of the method is to use a chain of two neural networks. The first network predicts the type of a primary particle for individual events, while the second infers the mass composition of an ensemble of events. We apply this method to the Monte-Carlo data for the Telescope Array Surface Detectors readings, on which it yields an unprecedented low error of 7\% for 4-component approximation. We also discuss the problems of applying the developed method to the experimental data, and the way they can be resolved.}

\maketitle

\section{Introduction}
\label{sec:introduction}

Ultra-high-energy cosmic rays (UHECR) are particles with energies greater than $10^{18}$~eV reaching Earth from the cosmos. Understanding their origin is one of the most important unresolved problems in modern astrophysics. The information on nuclear types of cosmic rays primary particles (proton, helium, etc.) is crucial for achieving progress in this task. However, identifying UHECRs mass composition, i.e. the types of primary particles and their relative fractions in the spectrum, remains a challenging task.

It is known that the UHECR flux consists mostly of protons and nuclei~\cite{kampert2012measurements} with only small admixtures of photons~\cite{TelescopeArray:2018rbt, PierreAuger:2016kuz} and neutrinos~\cite{PierreAuger:2019ens, TelescopeArray:2019mzl} allowed. UHECR mass composition is the key observable for distinguishing between various models of cosmic-ray production~\cite{Allard:2005cx, Aloisio:2007rc}, as well as understanding the physical nature~\cite{Zatsepin:1966jv, greisen1966end, Aloisio:2009sj} of their energy spectrum cutoff~\cite{HiRes:2007lra, PierreAuger:2008rol, TelescopeArray:2012qqu}. Knowing the mass composition is also important for the prospective cosmic ray astronomy. Namely, for proton-dominated composition UHECR arrival direction is expected to trace their sources. On the other hand, for nuclei-dominated composition this would not be the case due to large deflections in the galactic magnetic fields~\cite{Pshirkov:2013wka}.

The most accurate studies of UHECR mass composition are performed by measuring the longitudinal profiles of extended air showers by the fluorescence detectors~\cite{HiRes:2007lra, PierreAuger:2014gko, Abbasi:2014sfa, TelescopeArray:2018xyi, HiRes:2009fiy}. However, fluorescence detectors can operate only at moonless nights and under good weather conditions, which strongly limits the number of observable events per year.

An alternative approach is to use the readings of the surface detectors, which can collect data during almost the whole year. The trade-off is that surface detectors provide less informative data as compared to the one obtained with fluorescence detectors. Hence, to get a competitive accuracy, one needs to employ machine learning techniques. For example, this approach was undertaken in recent studies~\cite{TelescopeArray:2018bep, PierreAuger:2021fkf}.\footnote{We would also like to mention that it is possible to infer UHECR mass composition based on the detailed UHECR flux modeling and their energy spectrum~\cite{PierreAuger:2016use} or arrival directions~\cite{Kuznetsov:2020hso}.}

The aim of the present paper is to develop further machine learning techniques used for analyzing surface detectors data. We start by highlighting the two key aspects of the underlying problem.

The first one is related to the stochastic nature of an extensive air shower. Namely, extensive air showers are initiated by high-energy cosmic rays, which, upon entering Earth's atmosphere, interact with the nuclei of atoms. This results in an avalanche-like production of particles, whose footprints are measured by detectors on the ground level. The position of interaction points, as well as particles' properties after the interaction, follow some probability distributions. This makes the evolution of an air shower stochastic in nature and results in the variability of the footprints. In practice, the situation is complicated further by the presence of detectors triggered by occasional muons, irregular detector grid, saturated and offline detectors. As a consequence, different primary particles may yield similar footprints, which lowers the efficiency of standard machine learning techniques.

The second aspect is that there exist several hadronic interaction models used for simulating air shower evolution. Different models yield considerably different air shower properties at the ground level. Moreover, some experiments observe the excess of the observed muons signal as compared to the simulations \cite{abbasi2018study, aab2016testing, abu2000evidence, glushkov2008muon, dedenko2017testing}, while others do not \cite{apel2017probing,fomin2017no,gonzalez2015measuring,alekseev2021status} (see \cite{dembinski2019report} for a review). This imposes certain limitations on the use of hadronic interaction models for analyzing the actual experiment's data.

In the paper we develop a method, which is less affected by the first of the mentioned problems. The progress is achieved by introducing a chain of two neural networks for analyzing the data. The first network analyzes individual events and predicts the corresponding primary particle. The second network analyzes ensembles of events and reconstructs the mass composition based on the inference of the first network. We apply this method to the Monte-Carlo set with four cosmic ray primaries (proton, helium, nitrogen, and iron) for the surface detector readings of the Telescope Array experiment \cite{abu2012surface}. For a given hadronic model our approach yields unprecedentedly low error --- the fractions of primary particles in an ensemble of events are reconstructed with an error less than 7\%. 
	
The developed approach allows one to obtain accurate predictions for an ensemble of events despite high variability of data for individual events. This makes it applicable for a wide range of experiments with similar specifics of the data.

The paper is organized as follows. In section \ref{sec:TA} we briefly describe the Telescope Array experiment and the Monte-Carlo set we are using. Section \ref{sec:individual} is devoted to the discussion of the neural network used for analyzing individual events. In section \ref{sec:mass_spectrum} we introduce the second neural network, which is used for estimating UHECR mass composition. The issues that need to be resolved before applying our method to the actual experimental data, such as model dependence of the predictions, are discussed in section \ref{sec:mod_dep}. Finally, section \ref{sec:conclusion} concludes the paper.

\section{Deep learning for reconstructing mass composition}
\label{sec:DL}

\subsection{Data}
\label{sec:TA}

Telescope Array Observatory is located in Utah, USA, and is the largest cosmic ray detector in the Northern Hemisphere aimed at studying UHECR. The Telescope Array Surface Detector includes 507 scintillation surface detectors placed on a square grid, with a 1.2 km spacing, covering an area of approximately 700 km$^2$. Each detector has two layers of 1.2 cm thick plastic scintillator, which is sensitive to both electromagnetic and muon components of the air shower~\cite{abu2012surface}. Each layer measures time-resolved signal with 20 ns time bin size. The signal is converted to conventional units called ``minimal ionization particles'' (MIP) using the calibration information which is recorded every 10 minutes for each detector. The same calibration information is used in Monte-Carlo simulations as well.

An \textit{event} is recorded when the trigger condition is satisfied, see~\cite{abu2012surface} for details on the triggering system. For each event, the corresponding record includes \textit{i)} time series of the signal from each of the activated detectors and \textit{ii)} starting time of the recordings. Further, the parameters of the air shower, such as Linsley front curvature and energy of the primary particle, are estimated using the reconstruction procedure \cite{TelescopeArray:2012qqu} (see Appendix \ref{sec:B} for details).

Experimental data is modeled using standard TA procedure~\cite{TelescopeArray:2014nxa, TelescopeArray:2012qqu}: Monte-Carlo simulation using CORSIKA software \cite{heck1998corsika} with protons, helium, nitrogen, and iron as primary cosmic ray particles. High and low-energy hadronic interactions are modelled with QGSJET II-03 \cite{ostapchenko2006qgsjet} and FLUKA packages \citep{Ferrari:2005zk} correspondingly. The electromagnetic air shower component is simulated using EGS4 model~\citep{Nelson:1985ec}. The simulations use thinning and dethinning of extensive air showers~\cite{stokes2012dethinning} and GEANT4~\cite{agostinelli2003geant4} sampling library for producing detectors responses~\cite{ivanov2012energy}. Energy spectrum of the simulated events coincides with the one measured in the HiRes experiment~\cite{HiRes:2007lra}, and the number of simulated showers is the same for all primaries. The procedure accounts for offline detectors, saturated detectors and muons from randomly coincident low-energy showers.

To ensure good quality of the events, we are using data passing the following quality cuts: \textit{i)} energy of a primary particle estimated with the standard TA Surface Detector reconstruction procedure is larger than 1~EeV, \textit{ii)} reconstructed zenith angle of a primary particle is below 45$\degree$, \textit{iii)} the number of triggered detectors is 7 or more, \textit{iv)} small error of the reconstructed parameters (see Appendix \ref{sec:B} for details). In total, for all primaries we have 481 172 Monte-Carlo events satisfying these conditions.

The events were divided into training, validation, and test sets in proportion 8/1/1. Through out the paper, we use training and validation sets for training neural networks (validation set is used to invoke \textit{early stopping} to avoid overfitting). All of the metrics are evaluated on a test data set.

\subsection{Event-by-event analysis}
\label{sec:individual}

In previous work \cite{abbasi2019mass} the reconstructed parameters, such as Linsley front curvature and area-over-peak, were used to predict the mass of the primary particles using Boosted Decision Trees. As an improvement to this approach, we switch to using neural networks instead of Boosted Decision Trees for this task. This allows us to incorporate more data into the analysis (such as raw readings of the detectors) and employ advanced machine learning techniques. The benefits of using neural networks in similar tasks were demonstrated, for example, in \cite{ivanov2020using}, where they are used for reconstructing cosmic rays arrival directions.

One can pose the problem of identifying UHECR primary particle as a regression or classification tasks. This corresponds to predicting directly the mass of the primary particle or only its type (and thus the mass as well). Below we focus on the classification approach as it was found to result in slightly better overall performance. The regression approach is discussed in Appendix \ref{sec:A}.

\begin{figure}
\center{\includegraphics[width=0.98\linewidth]{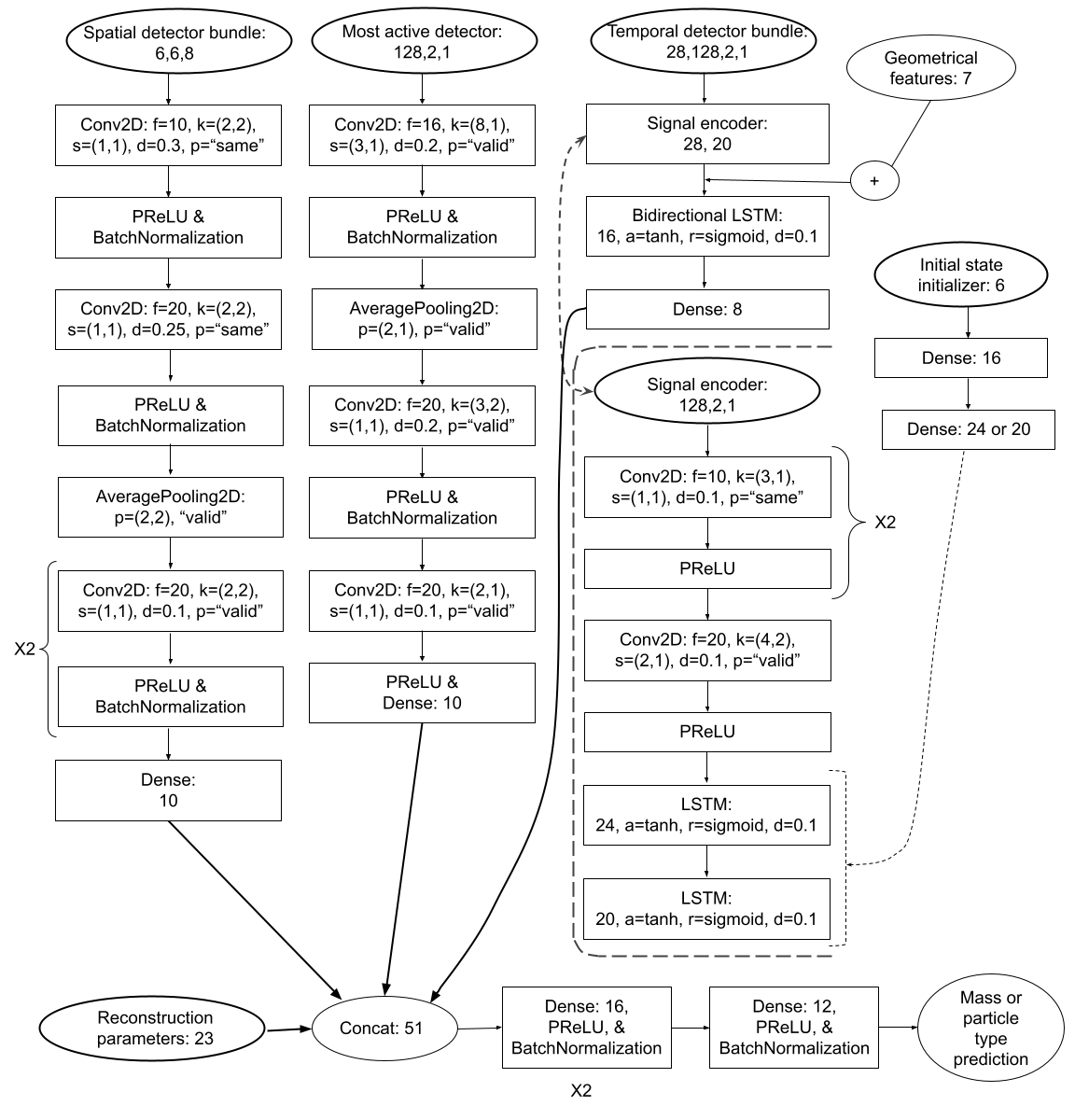}}\\
\caption{Architecture of the neural network. The flow of the data is shown by arrows. Layers names are given according to TensorFlow API.  Abbreviations: ``f'' --- number of filters, ``k'' --- kernel size, ``s'' --- strides, ``d'' --- dropout rate, ``p'' --- pooling size or padding regime, ``a'' --- activation function, ``r'' --- recurrent activation function, number after colon --- input shape or number of units.}
\label{classif_arch}
\end{figure}

The first step of our method consists of classifying individual events using a neural network. In what follows we refer to this network as \textit{the classifier}.It analyzes the raw data as registered by the surface detectors and high-level reconstruction parameters obtained with the standard TA reconstruction procedure \cite{TelescopeArray:2012qqu,Takeda:2002at,abu2013upper} (see also Appendix \ref{sec:B} for details). Classifier architecture, depicted in figure \ref{classif_arch}, can be split into four blocks, each of which is responsible for processing certain kind of data gathered from an event. Below we describe the blocks one by one, highlighting their general features. 

\textit{1. Spatial detectors bundle.} For each event we construct a six by six grid of detectors centered on the averaged position of the triggered detectors. In each ``detector cell'' we store eight parameters:
\begin{itemize}
    \item[1-2)] x and y coordinates of the detector relative to the reconstructed position of the air shower core,
    \item[3)] the position of the detector with respect to the conventional zero-ground level,
    \item[4)] detector's integral signal (sum of detector's readings),
    \item[5)] reconstructed time of the plane front arrival to the detector,
    \item[6)] time difference between the earliest timestamp of the recorded signal and the moment of the plane front arrival at the detector,
    \item[7)] whether detector was triggered,
    \item[8)] whether detector's readings were significantly longer than 128 bins (2.56 ms).
\end{itemize}

Altogether this can be treated as a six by six ``pixel image'' with 8 channels (instead of standard RGB). Convolution neural networks are known to be well-suited for analyzing such kind of data. Accordingly, we pass the above-mentioned data through a number of convolution layers to obtain spatial (geometrical) features of an event. 

\begin{figure}[h!]
\center{\includegraphics[width=0.5\linewidth]{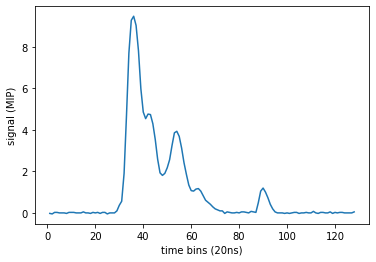}}\\
\caption{An example of detector's reading from one of the layers.}
\label{fig_waveform}
\end{figure}

\textit{2. Signal from the detector with the largest integral signal.} For a given detector the signal may last as long as 500 time bins. As in average the length of a signal is much shorter, we cut all signals to 128 bins to unify detectors' readings. It was verified that increasing this number does not improve neural network performance. A typical signal from one of the detector's layers is illustrated in figure \ref{fig_waveform}. Notice that the peaks in the signal are rather sharp. To soften them, we employ the signal conversion to the logarithmic scale by taking $ \ln(1+\text{signal}) $ as the input for the neural network. 

Detector's readings can be thought of as a one-dimensional image with two channels (one channel per detector layer). Correspondingly, we pass the signal through a series of convolution layers to obtain the features of the largest signal.\footnote{One can analyze the signal by recurrent layers. We tried LSTM and GRU cells but found no difference in the resulting precision, while convolutions are computationally much faster. } 

\textit{3. Temporal detector bundle.} To get temporal characteristics of an event, we collect all triggered detectors and order them in a one-dimensional chain according to the time of the plane front arrival. On Monte-Carlo data the number of the triggered detectors varies from 7 to 41 from event to event. Our practice showed that the chain can be safely cut down to 28 cells without losing neural network's performance, . In each cell we put the time series of the corresponding detector readings (reduced to 128 time bins as described above). 

To process the data, we first analyze the signal in each cell of the chain by a \textit{signal encoder}. As the first step, the encoder passes the time series of the signal through several convolution layers. Further, the resulting data is passed through recurrent cells, which are known to be well-suited for analyzing time-ordered information.\footnote{Unlike the analyzes of a signal from the detector with largest signal, the use of recurrent cells yields an improvement in the performance.} This procedure yields signal encodings for each of the detectors.

Similarly to \cite{aab2021extraction}, we use a small fully connected neural network for setting the initial state of the recurrent cells analyzing the signal in the signal encoder. The input data for this network includes 6 parameters: x, y, and z coordinates of the detector, reconstructed time of the plane front arrival, time difference between starting point of detectors' recordings and the time of plane front arrival, and reconstructed zenith angle of the primary particle. 

As the next step, the obtained encodings are concatenated with geometrical characteristics of the corresponding detectors (parameters 1--6 and 8 used in spatial detector bundle). Together they represent air shower and detector properties at detector's location, ordered according to the time of the plane front arrival. Finally, the 28 cells with thus obtained information are analyzed by recurrent layers to obtain temporal features of an event. 

\textit{4. Combination layer.} At the final layer we concatenate all of the obtained features and supplement them with useful reconstruction parameters. The latter include zenith angle, normalization of the shower lateral distribution profile $S_{800}$, and composition-sensitive parameters (see  Appendix \ref{sec:B} for the full list of parameters). Together they are passed through several fully-connected layers to make the final prediction. The latter can be either the mass of the primary particle or its type.

The most important block in the neural network is the temporal detector bundle as it utilizes all available information on an event. Reconstruction parameters carry high-level information, like air shower's front curvature, and thus are expected to be the next by relevance. Spatial detector bundle yields geometrical features of an event, placing it third by relevance. Finally, the block analyzing the signal from the most active detector is of least importance as it utilizes the least amount of information. We confirm this ranking of the blocks by providing their individual accuracies later in this section.

The classifier is implemented in TensorFlow \cite{abadi2016tensorflow} and optimized using Keras-Tuner~\cite{omalley2019kerastuner} on 1/3 of the training data. The full architecture of the model, including layer specifics, is presented in figure \ref{classif_arch}. In total, the neural network has $\sim 35~ 000 $ trainable parameters. For training the network we used Adam optimizer \cite{kingma2014adam} with \textit{amsgrad} modification \cite{reddi2019convergence}.

We found it beneficial to train the classifier in two steps. First, the network is trained on all available events. Then, at the second stage, its training is continued only on the events in a certain energy bin(the energy is estimated via the reconstruction procedure). This approach allows the network first to capture general features of the events, and then fine-tune itself for specific energies.

To illustrate the method, we further consider the network trained for the events in the $ 10^{18.75}$--$10^{19.00} $~eV energy bin during the second stage of training. The corresponding Monte-Carlo set consists of 103 752 events, which were split into training, validation, and test sets in proportion 8/1/1. The results for other energy bins are similar and are presented in the table at the end of next section.

On an event-by-event basis, the accuracy of our best-trained classifier was 40.2\%. This is far from being an ideal accuracy, yet much better than random guessing (25\%) and the accuracy of boosted decision trees used earlier for this problem ($\sim$31\%). We have also established the contribution of each of the neural network's blocks to the final accuracy by training neural networks consisting of only one of the blocks. The results are presented in table \ref{nn_parts_acc}, and they confirm our expectations on the importance of the blocks.

\begin{table}
\centering
\begin{tabularx}{0.98\textwidth}{ 
  | >{\centering\arraybackslash}X | >{\centering\arraybackslash}X 
  | >{\centering\arraybackslash}X | >{\centering\arraybackslash}X | >{\centering\arraybackslash}X | }
\hline
 & \small{temporal detector bundle} & \small{reconstruction parameters} & \small{spatial detector bundle} & \small{most active detector} \\ [0.5ex]
\hline 
\small{accuracy} & 39.5\% & 33.8\% & 32.9\% & 31.0\% \\ 
\hline 
\end{tabularx}
\caption{Accuracies of different blocks of the classifier.}
\label{nn_parts_acc}
\end{table}

\subsection{Identifying mass composition}
\label{sec:mass_spectrum}

A naive method for obtaining the mass spectrum for an ensemble of events would be to average the predictions of the classifier over events.
However, due to low accuracy of the classifier, this would yield large errors. To improve the quality of predictions, one can employ the techniques, which make use of the available statistics of the events.

One of such techniques employs the confusion matrix and goes as follows. Let $ y_i $ be the averaged (over events) predictions of the classifier for all samples of $i$-th class (particle type). Then the confusion matrix $ C $ is defined by the following equation:
\begin{equation}
y_i = C \hat{y}_i \,, ~~~ \forall i \;.
\end{equation}      
where $ \hat{y}_i $ are the correct predictions (all zeros except for $1$ at $i$-th position). Usually, and as we do in the paper, the confusion matrix is evaluated using the validation data set. Then, given an ensemble of events of interest, one can improve the averaged predictions of the classifier by multiplying them with $C^{-1}$. 

As an enhancement to the described method, instead of employing the confusion matrix we suggest using a new neural network for improving averaged predictions of the classifier. This yields two benefits. First, the transformation of the averaged predictions becomes non-linear, hence allowing for more precise mappings. Second, the neural network can take into account not only the averaged predictions, but higher-order moments of the distribution as well. This allows to improve the predictions by utilizing more information. Below we refer to this additional neural network as \textit{the converter}. 

The converter's architecture consists of 3 hidden fully-connected dense layers, having 32, 16, and 8 units respectively, with PReLU activation function \cite{he2015delving}. The last layer is a dense layer with 4 units and softmax activation function. We observed that introducing more layers or increasing the number of units in the network does not improve the performance.

For training the converter, we made 10 000 ensembles of events, consisting of 5 000 samples each, from all available events in a given energy bin. To sample them, we first choose at random the fractions of four primary particles in an ensemble, and then randomly sample events for each of the particles. The ensembles were divided into training, validation, and test sets in proportion 8/1/1. Note that, as we have a limited number of events, some of the ensembles have significant intersections. Moreover, by sampling from all events, we have mixed classifier's training, validation, and test sets. One may worry that this will lead to overfitting. By reserving classifier's test set for making test ensembles only, we verified that this is not the case. Note also that as the converter has few trainable parameters, it is expected to be robust to overfitting.  

All of the ensembles were passed through the classifier. For a given ensemble, this yields 5000$\times$4 predictions, corresponding to the classifier confidence on what was the primary particle in each of the events. By averaging over the events we get 4 numbers --- the classifier prediction of the fractions of the elements in the ensemble. We also obtain the standard deviations of the predictions among the events, representing the diversity of the events in the ensemble. Together, these 8 numbers were given to the converter, which was trained to yield the true fractions of elements in the ensemble. It was observed that employing higher order moments of the distribution does not improve converter's performance.

\begin{figure}
     \centering
     \begin{subfigure}[b]{0.45\textwidth}
         \centering
         \includegraphics[width=\textwidth]{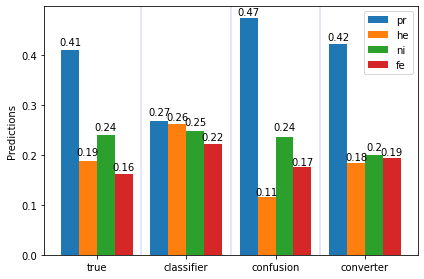}
         \caption{}
     \end{subfigure}
     \hfill
     \begin{subfigure}[b]{0.45\textwidth}
         \centering
         \includegraphics[width=\textwidth]{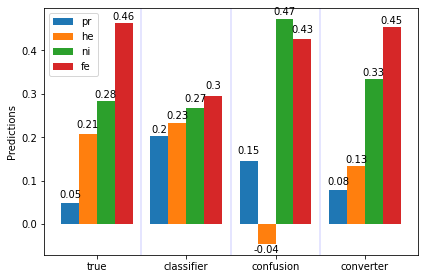}
         \caption{}
     \end{subfigure}
        \caption{Results of predicting the ensemble composition for two different mixes. The fractions of elements are marked on the vertical axis. In each figure the four histograms, from left to right, correspond to 1) true fractions of elements, 2) averaged classifier predictions, 3) confusion matrix-corrected predictions, and 4) predictions of the converter.}
        \label{methods_comparison}
\end{figure}

Typical results of applying the chain of two neural networks for estimating the fractions of elements in an ensemble are depicted in figure \ref{methods_comparison}. The averaged predictions of the classifier almost always indicate equal fractions of elements and thus are unreliable. The confusion matrix approach performs better, but sometimes yields significant reconstruction errors. Finally, the converter shows the best results. We also observed that the accuracy of the final predictions grows rapidly with the precision of the classifier. This is the reason why we have not dropped any blocks from the classifier architecture.

To estimate the error of the method, for each of the primary particles we evaluated the mean absolute error (MAE) between its true fraction in ensembles and the predicted one. For this purpose we used 2 000 ensembles from the validation set, the results are presented in table \ref{errors_onebin}. As it can be expected, the method with the converter yields best performance. We estimate the absolute error of our method as $ 7 $\% --- the largest MAE between true and reconstructed fractions among elements.

An alternative metric that can be used for evaluating the performances of the methods is the maximum absolute error between true fractions and predicted ones. For the predictions obtained with the classifier, confusion matrix, and converter the corresponding errors are 0.6, 0,3, and 0.1 correspondingly. This shows that the converter yields best results.

\begin{table}
\centering
\begin{tabularx}{0.7\textwidth}{ 
  | >{\centering\arraybackslash}X | >{\centering\arraybackslash}X 
  | >{\centering\arraybackslash}X | >{\centering\arraybackslash}X | >{\centering\arraybackslash}X | }
\hline
 & proton & helium & nitrogen & iron \\ [0.5ex]
\hline 
classifier & 0.10 & 0.11 & 0.11 & 0.09 \\ 
\hline
confusion & 0.06 & 0.14 & 0.12 & 0.04 \\
\hline
converter & 0.03 & 0.07 & 0.06 & 0.02 \\ [1ex] 
\hline
\end{tabularx}
\caption{MAE between true and reconstructed fractions of elements for different methods and particles.}
\label{errors_onebin}
\end{table}

If Monte-Carlo set includes all types of primary particles present in the data, the predictions of the converter can be directly interpreted as fractions of elements in an ensemble. In the opposite case one should use the prediction as weights for estimating the mean mass, $ m $, of primary particles:
\begin{equation}
m = \sum_i p_i m_i \;,
\end{equation} 
where $ p_i $ and $ m_i $ are, respectively, the predictions and masses of the elements.

Errors for reconstructing the mass composition using the converter in other energy bins are summarizes in table \ref{errors_all}. Reconstruction accuracy of the method grows with the energy. This is due to the fact that the higher the energy of the primary particle is, the better is the quality of its footprint on the detectors. Correspondingly, the classifier, and hence the converter, can better distinguish different primary particles. 

\begin{table} [h]
\centering
\begin{tabularx}{0.95\textwidth}{ 
  | >{\centering\arraybackslash}p{3cm} | >{\centering\arraybackslash}X | >{\centering\arraybackslash}X 
  | >{\centering\arraybackslash}X | >{\centering\arraybackslash}X | >{\centering\arraybackslash}X | }
\hline
\small{energy bin, log scale} & \small{18--18.25} & \small{18.25--18}.5 & \small{18.5--18.75} & \small{18.75--19} & \small{$>19$} \\ [0.5ex]
\hline 
\small{averaged MAE} & \small{0.072} & \small{0.053} & \small{0.048} & \small{0.043} & \small{0.038} \\ 
\hline
\end{tabularx}
\caption{MAE between true and reconstructed fractions of elements, averaged over the elements for different energy bins. 
\label{errors_all}}
\end{table}

\section{Discussion}
\label{sec:mod_dep}

Monte-Carlo simulations of UHECR are known to be in disagreement with the data measured in various experiments. Namely, a number of collaborations reported on the excess of registered number of muons on the ground level as compared to the simulations \cite{abbasi2018study,bellido2018muon,cazon2020working,aab2015muons,aab2016testing}. This motivated the refinement of high-energy hadronic interaction models used for simulations by incorporating the cross-sections measured at the LHC. For some of the experiment, this has resolved the issue, while for others it hasn't \cite{cazon2020working}. Moreover, a recent study \cite{abbasi2022density} has shown that while such models improve the agreement at high energies, they produce too many muons at lower energies.

There are ongoing researches on the development and identification of a reliable hadronic interaction model. For example, the simulated and measured profiles of extensive air showers are compared in \cite{aab2016testing, apel2017probing}, and the recorded muon component of the signal are compared in \cite{abbasi2018study, aab2021extraction, dedenko2017testing}. However, no universal and reliable hadronic interaction model has been established so far. For the TA experiment, the reference high-energy interaction model was chosen to be QGSJET II-03. This is the main reason why we used this model in our analysis. The developed method, which is the main contribution of this paper, can be applied to data simulated with any hadronic interaction model.

\begin{figure}
     \centering
     \begin{subfigure}[b]{0.45\textwidth}
         \centering
         \includegraphics[width=\textwidth]{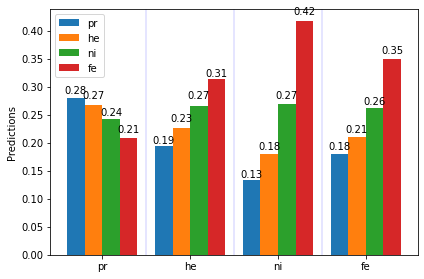}
         \caption{}
         \label{qgs4_averaged}
     \end{subfigure}
     \hfill
     \begin{subfigure}[b]{0.45\textwidth}
         \centering
         \includegraphics[width=\textwidth]{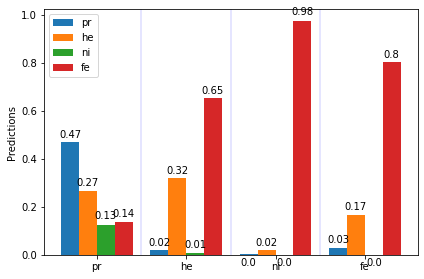}
         \caption{}
         \label{qgs4_converter}
     \end{subfigure}
        \caption{ Predictions of the classifier and converter on the data simulated with QGSJET II-04 hadronic model. (a) --- averaged predictions of the classifier, (b) --- converter predictions. Four histograms on each of the panels, from left to right, show the predictions of fractions of elements in ensembles consisting only of protons, helium, nitrogen, and iron respectively.}
        \label{qgs3_on_qgs4}
\end{figure}

To study the error of our method associated with the discrepancy between the simulation and real world evolution of air showers, we have done the following. First, we trained the classifier and the converter on the data for QGSJET~II-03 model only. Further, using the data simulated with QGSJET~II-04 hadronic model, we made ``pure'' ensembles, consisting only of protons, helium, nitrogen, and iron respectively. These pure ensembles were analyzed by the classifier and then by the converter. The predictions are shown in figure \ref{qgs3_on_qgs4}.

From panels \ref{qgs4_averaged} and \ref{qgs4_converter} one concludes that the classifier strongly overestimates masses of primary particles. Interestingly, the misclassification rates are larger for more massive primary particles.\footnote{Predictions for the iron are an exception, since we do not have heavier elements in our training data.} This large systematic model error indicates that our neural network is sensitive to the features of a given hadronic interaction model. Therefore, before applying the method to the actual experimental data, it should be made more robust. 

\begin{table} [ht]
\centering
\begin{tabularx}{0.95\textwidth}{ 
  | >{\centering\arraybackslash}p{3cm} | >{\centering\arraybackslash}X | >{\centering\arraybackslash}X 
  | >{\centering\arraybackslash}X | >{\centering\arraybackslash}X | >{\centering\arraybackslash}X | }
\hline
primary particle & proton & helium & nitrogen & iron & averaged \\ [0.5ex]
\hline 
reconstruction error & 0.53 & 0.68 & 1.00 & 0.20 & 0.60  \\ 
\hline
\end{tabularx}
\caption{Systematic model error of the method, estimated as the fraction of misclassified elements.}
\label{sys_err}
\end{table}

To reduce the model choice systematic error, one may try to train the classifier on a Monte-Carlo simulations for various hadronic interaction models. Presumably, this will force the neural network to pick up only model-independent features, and thus reach model-independence of the predictions.

We tried the second approach by combining Monte-Carlo simulated events for both QGSJET II-03 and QGSJET II-04 hadronic models. When the classifier was trained and tested on each of the models separately, it reached the accuracies 40.2\% and 42.3\% respectively. Interestingly, this shows that QGSJET II-04 hadronic model produces more distinguishable events. The classifier trained and tested on the combined data yields 41.3\% accuracy, which is close to the arithmetic mean of the models trained on each set of events separately. This indicates that model-independent training is possible, and to fully implement it one needs enough data simulated with more hadronic interaction models. Full implementation of this approach, including tests with other hadronic interactions models, will be the subject of our next research.

\section{Conclusion}
\label{sec:conclusion}

We have developed a method that allows one to achieve high accuracy in analyzing ensembles of events despite high variability of the data for individual events. The key idea of the method is to use a chain of two neural networks --- one for analyzing individual events and one for analyzing the ensembles of events.

The method has been applied for identifying the mass composition of ultra-high-energy cosmic rays on the example of Monte-Carlo data for the Telescope Array Surface Detector. On this task, our approach yielded an unprecedented low error of 7\% in reconstructing the fractions of 4 primary particles in ensembles of events. We have discussed the systematic error due to the choice of the hadronic interaction model and highlighted possible ways of how this error can be reduced.  

\acknowledgments

The work is supported by the Ministry of Education of Russian Federation, grant number 075-15-2020-778.

\appendix

\section{Regression approach}
\label{sec:A}

Instead of predicting types of primary particle for individual events, one may directly estimate their masses, i.e. consider the task as the regression problem. The benefit of choosing this approach is that, if it is precise enough, it would allow one to identify in an experimental data a primary particle that was not present in the Monte-Carlo set. Below we study this approach.   

It is known that air shower parameters are most conveniently expressed as a function of the logarithm of the primary particle mass. Hence adopting the logarithm mass scale would make it easier for the neural network to learn the corresponding dependence. We employed this parametrization and labeled the particles with logarithms of their masses. For simplicity, the labels for protons, helium, nitrogen, and iron were approximated as 0, 1, 2, and 3 respectively (logarithm in base 4). 

For training the network, we took the loss function to be the mean squared error, used Adam optimizer, and employed early stopping to avoid overfitting. Below we refer to this network as the \textit{regressor}.

\begin{figure}
     \centering
     \begin{subfigure}[b]{0.48\textwidth}
         \centering
         \includegraphics[width=\textwidth]{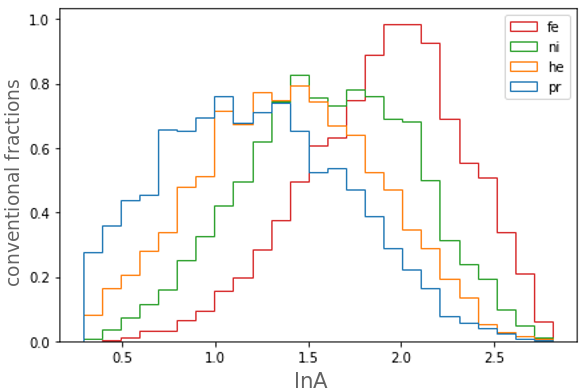}
         \caption{}
     \end{subfigure}
     \hfill
     \begin{subfigure}[b]{0.48\textwidth}
         \centering
         \includegraphics[width=\textwidth]{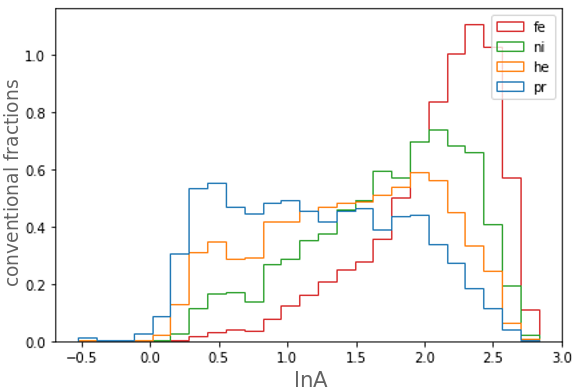}
         \caption{}
     \end{subfigure}
        \caption{Histograms for predictions on the validation set. Case a) shows predictions for the neural network trained on all available primaries, while in case b) helium was excluded from the training data.}
\label{regression_2_models}
\end{figure}

To study sensitivity of the network to nuclei not present in the data set, we trained two models. The first model was trained on all available data, while for the second we excluded all events with helium as a primary particle. The histograms illustrating the performances of the models are depicted in figure \ref{regression_2_models}.

The figure shows that the regressor performs poorly on separating the elements. Namely, the histograms for different elements have large overlapping regions, and, for a given element, the averaged prediction of the logarithmic mass is far from its true value. Even for the model trained on all available data, the averaged predictions for protons, helium, nitrogen, and iron are 1.32, 1.40, 1.61, and 1.90 correspondingly. A positive observation is that the model that has never seen helium places it correctly somewhere between proton and nitrogen, with a similar shape of the distribution.

To obtain the mass composition, one can introduce the second neural network in a way, similar to how it was done in the classification regime.  Namely, given an ensemble of events, one can obtain the distribution of the predictions of the regressor, similar to fig. \ref{regression_2_models}. Further, this distribution can be split into $ n $ bins, thus forming a one-dimensional n-pixel ``black and white'' (as there is only one channel --- value of the distribution at the given bin) image. The second neural network (the converter) in this case would be a convolutional network, analyzing this image and yielding true fractions of elements in the ensemble as the output.

We have tried this approach, with the predictions of the regressor split into 50 bins. It was observed, however, that this yields a larger error, approximately $ 0.5\% $ bigger, than in the classification regime.

\section{Composition-sensitive parameters and cuts}
\label{sec:B}

The composition-sensitive parameters are obtained in a two-step procedure. At the first step a joint fit of the shower geometry and lateral distribution function (LDF) is performed using each detector's trigger time and total signal. The shower front curvature parameter is determined at this fit. At the second step a number of additional parameters are estimated using time-resolved signals at each triggered detector. The details and corresponding references are given below.

The joint geometry and LDF fit has 7 free parameters: 
\begin{itemize}
    \item[1-2)] Position of the shower core, $ \vec{r}_{core} = (x_{core},~y_{core},~0) $, defined in the coordinate system centered at the Central Laser Facility.
    \item[3-4)] Arrival direction of the shower given by zenith and azimuthal angles, $ \theta $ and $ \phi $.
    \item [5)] Normalization of the shower's lateral distribution profile, $ S_{800} $.
    \item[6)] The time offset, $ t_0 $.
    \item[7)] Linsley front curvature parameter, $ a $. 
\end{itemize}
The target functions for the fit are
\begin{subequations} 
\begin{gather}
 t(\vec{r}) = t_0 + t_{plane}(\vec{r}) + a\times LDF(r)\times\left(1+\frac{r}{R_l}\right)^{1.5} \;, \\
 S(r) = S_{800}\times LDF(r) \;,
\end{gather}
\end{subequations}
where $ r $ is the distance from the detector to the shower core (the shortest distance between a point and a line), $ t_{plane}(\vec{r}) $ is the arrival time of the shower plane to the detector located at $ \vec{r} $, and $ LDF(r) $ is the empirical lateral distribution profile, introduced in AGASA experiment \cite{teshima1986properties} and then modified for TA data~\cite{TelescopeArray:2012qqu}:
\begin{subequations} 
\begin{gather}
 t_{plane}(\vec{r}) = \frac{1}{c}\times (\vec{n},(\vec{r}-\vec{r}_{core}))  \;, \\
 LDF(r) = f(r)/f(800m) \;, \\
 f(r) = \left(\frac{r}{R_m}\right)^{-1.2} \left(1+\frac{r}{R_m}\right)^{-(\eta-1.2)} \left(1+\frac{r^2}{R_1^2}\right)^{-0.5} \;.
\end{gather}
\end{subequations}
Here $ c $ is the speed of light, $ \vec{n}=\vec{n}(\theta,\phi) $ is the unit vector along the direction of the shower axis, $ (\cdot,\cdot) $ stands for the scalar product, and
\begin{subequations} 
\begin{gather}
R_m = 90~m \,, ~~~ R_1 = 1000~m \,, ~~~ R_L=30~m \;, \\
\eta = 3.97 - 1.79(\text{sec}(\theta)-1) \;.
\end{gather}
\end{subequations}

The energy of the primary particle is estimated as a function of $S_{800}$ and $\theta$ with the lookup table obtained from the Monte-Carlo simulations~\cite{2014arXiv1403.0644T,TelescopeArray:2012qqu}. 

Below is the complete list of the 23 composition-sensitive parameters used in the present paper. These parameters generally resemble ones from~\cite{abbasi2019mass}.

\begin{itemize}
    \item[1-3)] Reconstructed Linsley front curvature parameter ($a$), zenith angle ($ \theta $), and signal density at 800 m from the shower core ($ S_{800} $), as described above.
    \item[4-5)] $ \chi^2 $ and $ d.o.f. $ for the joint geometry and LDF fit.
    \item[6-7)] Area-over-peak of the signal at 1200 m and the corresponding slope parameter \cite{abraham2008upper}.
    \item[8)] Number of triggered detectors.
    \item[9-10)] Number of detectors that were used and excluded\footnote{Some of the detectors may be triggered by particles unrelated to the air shower. They are identified and excluded from the fit by the fitting procedure.} from the geometry fit \cite{TelescopeArray:2012qqu}.
    \item[11-15)] $ S_b \equiv \sum_i \left( S_i\times\left(\frac{r_i}{r_0}\right)^b \right) $ for five values of $b$: 2.5, 3.0, 3.5, 4.0, and 4.5. Here $ S_i $ and $ r_i $ are, respectively, the integral signal of $i$-th detector and its distance to the shower core, and $ r_0 = 1200~m $~\cite{Ros:2011zg}.
    \item[16)] The sum of integral signals from all of the detectors.
    \item[17)] Asymmetry of the signal at the upper and lower layers of detectors \cite{abbasi2019mass}.
    \item[18)] Total number of peaks within all detectors \cite{abbasi2019mass}.
    \item[19)] Number of peaks for the detector with the largest signal.
    \item[20-21)] Number of peaks present in the upper layer and not in the lower, and vice versa.
    \item[22)] Minimal distance between the reconstructed air shower core and detectors array edge.
    \item[23)] Number of detectors with signal longer than 128 time bins.
\end{itemize}

We have examined the usefulness of these parameters by training a neural network with two hidden fully-connected (dense) layers. The input of the network was one of the composition-sensitive parameters, and the output --- the predicted type of the particle. The resulting ranking is as follows (random guessing corresponds to 25\% accuracy):

\begin{itemize}
    \item $ \sim 30\% $ accuracy: Linsley front curvature.
    \item 27\%--29\% accuracy:  total number of triggered detectors, number of detectors used for the geometry fit, reconstructed zenith angle, $ d.o.f $, AOP slope, $ S_{4.5} $, and total number of peaks.
    \item $ \sim 26\% $ accuracy: $ S_{800} $, $ \chi^2 $, $ S_{3.0} $, $ S_{3.5} $, $ S_{4.0} $, number of detectors excluded from the geometry fit, asymmetry of the signal in the upper and lower detectors, and total number of peaks present in the upper detector's layer and not in the lower.
    \item less than 26\% accuracy: all others.
\end{itemize}

To ensure good quality of the events the neural network is trained on, we impose the following quality cuts:
\begin{itemize}
\item At least 7 detectors were triggered.
\item Zenith angle is below 45\degree .
\item The distance between the reconstructed air shower core and detectors array edge is more than 1200 m.
\item $ \chi^2 / d.o.f. $ is less than 4 for separate geometry and the LDF fits, and is less than 5 for their joint fit.
\item The error of estimating the arrival direction is less than 5\degree.
\item The error of estimating $ S_{800} $ is less than 25\%.
\end{itemize}

\bibliography{ref_TAspectrum}

\end{document}